\documentclass[preprint,aps,showpacs,showkeys,amsmath,amssymb,nofootinbib]{revtex4}
\usepackage{graphicx}% Include figure files
\usepackage{dcolumn}% Align table columns on decimal point
\usepackage{bm}% bold math

\def\bge{\begin{equation}}
\def\ene{\end{equation}}
\def\bg{\begin{eqnarray}}
\def\en{\end{eqnarray}}

\begin{document}

\preprint{JLAB-THY-08-807}

\title{Quark-meson coupling model with the cloudy bag}% Force line breaks with \\

\author{S. Nagai}
%\altaffiliation[Also at ]{
%}%Lines break automatically or can be forced with \\
\author{T. Miyatsu}
%\altaffiliation[Also at ]
%}%Lines break automatically or can be forced with \\
\author{K. Saito}%
\email{ksaito@ph.noda.tus.ac.jp}
\affiliation{%
Department of Physics, Faculty of Science and Technology,\\
Tokyo University of Science, Noda 278-8510, Japan 
}%

\author{K. Tsushima}
%\altaffiliation
%[Also at ]
\affiliation{Excited Baryon Analysis Center (EBAC), Theory Group, 
Thomas Jefferson National Accelerator Facility\\
12000 Jefferson Avenue, Newport News, VA 23606, USA% with \\
}%

\date{\today}% It is always \today, today,

\begin{abstract}
Using the volume coupling version of the cloudy bag model, the quark-meson coupling model is extended to study 
the role of pion field and the properties of nuclear matter. 
The extended model includes the effect of gluon exchange as well as 
the pion-cloud effect,  
and provides a good description of the nuclear matter properties. 
The relationship between the extended model and the EFT approach 
to nuclear matter is also discussed. 
\end{abstract}

%Use showkeys class option if keyword
\maketitle

The quark-meson coupling (QMC) model~\cite{qmc} can be considered as an extension of Quantum Hadrodynamics (QHD) 
to include the effect of the internal structure of a nucleon in matter.  The model describes a nuclear system by non-overlapping 
MIT bags, in which the confined quarks interact through the self-consistent exchange of isoscalar, scalar ($\sigma$) and vector 
($\omega$) mesons.  In the past few decades, it has been extensively 
developed and applied to various nuclear phenomena with 
tremendous success~\cite{qmc}. 

On the other hand, a major breakthrough occurred in the problem of nucleon-nucleon (NN) force by introducing the concept of an effective 
field theory (EFT)~\cite{mach}. 
The QCD Lagrangian for massless up and down quarks is chirally symmetric, and the axial symmetry is spontaneously broken. 
This implies the existence of the massless Nambu-Goldstone bosons, namely the pions. The non-zero pion mass is then a consequence of
the fact that the light quark has a small mass.
Thus, one arrives at a low-energy scenario that pions and nucleons 
(and possibly deltas) interact via a force
governed by spontaneously broken, approximate chiral symmetry. 

In EFT, the degrees of freedom of quarks and gluons (including heavy mesons and nucleon resonances) should be 
{\it integrated out} (or {\it cutoff}), because a probe with wavelength $\lambda$ is insensitive 
to details of structure at distances much smaller than it~\cite{lepage}.  
Instead, it is necessary to add local (contact) interactions with low-energy constants (LECs) to the Lagrangian 
to mimic the effect of the true short distance physics. 
The LECs are determined empirically from fits to $\pi$N and/or NN scattering data, and vary with the momentum cutoff ($\sim \lambda^{-1}$) 
accounting for quantum fluctuations excluded by the cutoff~\cite{lepage}.  
Up to N$^3$LO, EFT can provide the peripheral NN scattering data (below about $250$ MeV lab. energy) very accurately~\cite{mach2}. 
 
Recently, EFT has been intensively applied to the problem of nuclear matter. 
In addition to the usual (small momentum) expansion in the free NN or $\pi$N scattering, physical observables in matter are 
expanded in terms of the Fermi momentum $k_F$, which is also a relevant, small scale. 
Such density dependence arises from the 
Pauli blocking effect in matter, i.e., the medium insertion including the step-function, $\theta(k_F - |{\vec p}\,|)$, in the nucleon 
propagator. Then, the strengths of the LECs are fine-tuned so as to reproduce the nuclear matter properties~\cite{density}. 
%, which may be different from the values determined in the EFT treatment of (free) $NN$ or $\pi N$ 
%scattering~\cite{density}. 

%Now, is it {\it unnecessary} to consider the degrees of freedom of quarks and gluons in low-energy nuclear systems? 
If the internal structure of the nucleon were completely {\it frozen} in a nuclear medium or the same as that in free space, 
it might be sufficient to consider the density dependence solely 
stemming from the Pauli blocking effect. 
However, if the in-medium nucleon were metamorphosed depending on 
the nuclear density $\rho_B$, 
the situation may be different.  In fact, the evidence for 
the medium modification of nucleon 
structure was observed in polarization transfer measurement 
in the quasi-elastic ($e$, $e^\prime p$) reaction at the 
Thomas Jefferson national accelerator facility, and the result supports the prediction of the QMC model~\cite{jlab}. 
It also seems vital to consider the internal structure change of the nucleon to understand the nuclear EMC effect~\cite{emc}.  

The QMC model can describe the medium modification of the nucleon structure through the quark model, and 
%(although it is rather model-dependent), and  
predict the density (or mean scalar-field) dependence of physical quantities~\cite{qmc}. 
Expanding such modification in terms of $k_F$ and comparing with 
the values of LECs given in the EFT approach,\footnote{
In fact, from the poin of view of the quark and gluon degrees of freedom, the QMC model can explain the values of 
the coefficients appearing in the familiar (contact) Skyrme force in conventional nuclear physics~\cite{skyrme}.
}  
it may be possible to study whether the internal structure change of a nucleon 
indeed shows up in matter, 
since the LECs involve all information on the short distance physics. 

%However, it may be a very complex project. As the first step 
To carry out such a complicated investigation, 
as a first step, we need to develop  
a new version of the QMC model for nuclear matter, 
where the structure of nucleon (and delta) 
is treated based on chiral symmetry. 
We attempt this in the present study using the 
volume coupling version of the cloudy bag model (CBM), which incorporates 
major results of the current algebra for 
low energy $\pi$N scattering~\cite{tony}.  

The Lagrangian density for the volume coupling version of the CBM in flavor SU(2) is given by~\cite{tony} 
\bge
{\cal L}_{CBM} =  \left[ {\bar \psi} \left\{ i\gamma_\mu {\cal D}^\mu + \frac{1}{2f_\pi} \gamma_\mu \gamma_5 
{\vec \tau} 
\cdot (D^\mu {\vec \phi}) \right\} \psi -B \right] \theta_V - \frac{1}{2}{\bar \psi} \psi \delta_S 
+ \frac{1}{2} ( D_\mu {\vec \phi} \,)^2 + {\cal L}_{\chi B} , \label{chiralv-lag}
\ene
with $\psi$ the quark field, ${\vec \phi}$ the pion field, 
$\phi = ({\vec \phi}\cdot {\vec \phi})^{1/2}$, ${\hat \phi} = {\vec \phi}/|{\vec \phi}|$, $f_\pi$ ($=93$ MeV) the pion decay constant, 
$B$ the bag constant, $\theta_V$ the step function for the bag, $\delta_S$ the surface $\delta$-function, 
$D_\mu {\vec \phi} = (\partial_\mu \phi) {\hat \phi} + f_\pi \sin(\phi /f_\pi)\partial_\mu {\hat \phi}$ and 
${\cal D}^\mu \psi = \partial^\mu \psi - \frac{i}{2}[ \cos (\phi /f_\pi)-1 ] {\vec \tau}\cdot ({\hat \phi} 
\times \partial^\mu {\hat \phi}) \psi$.
The last term includes the quark mass, $m$, which explicitly breaks chiral symmetry, and 
the pion mass, $m_\pi (=138$ MeV): 
${\cal L}_{\chi B} = - m {\bar \psi} e^{-i{\vec \tau}\cdot {\vec \phi}\gamma_5 /f_\pi} \psi \theta_V - \frac{1}{2}m_\pi^2 
{\vec \phi}\,^2$.

As in the CBM, we linearize the pion field and keep ${\cal O}(1/f_\pi)$ 
(the convergence properties of the CBM were given in Ref.~\cite{tony}).  
The Lagrangian density then reads 
\bg
{\cal L}_{CBM} &=& 
\left[ {\bar \psi} \left\{ i\gamma_\mu \partial^\mu -m + i \frac{m}{f_\pi}\gamma_5 {\vec \tau} \cdot {\vec \phi} 
+ \frac{1}{2f_\pi} \gamma_\mu \gamma_5 {\vec \tau} \cdot (\partial^\mu {\vec \phi}) \right\} \psi -B \right] \theta_V 
- \frac{1}{2}{\bar \psi} \psi \delta_S  \nonumber \\
&+& \frac{1}{2} ( \partial_\mu {\vec \phi}\, )^2 - \frac{1}{2}m_\pi^2 {\vec \phi}\,^2 . \label{cbm-lag}
\en
Here the pion field interacts with the quark through both the pseudovector (pv) and 
pseudoscalar (ps) couplings.  The strength of the ps coupling is 
${\cal O}(m/f_\pi)$, which explicitly shows the breaking scale of chiral symmetry.  
%Thus, the pseudoscalar coupling may provide a measure of the chiral symmetry breaking. 

We introduce the gluon field as well. The resulting Lagrangian density is thus given by 
\bge
{\cal L}_{CBM} = {\cal L}_{BAG} + {\cal L}_{\pi} + {\cal L}_{g} + {\cal L}_{int} , \label{cbm-lag2}
\ene
where 
\bge
{\cal L}_{BAG} = \left[ {\bar \psi} ( i\gamma_\mu \partial^\mu -m) \psi -B \right] \theta_V - \frac{1}{2}{\bar \psi} \psi 
\delta_S , 
\ene
\bge
{\cal L}_{int} = {\bar \psi} \left[ i \frac{m}{f_\pi}\gamma_5 {\vec \tau} \cdot {\vec \phi} 
+ \frac{1}{2f_\pi} \gamma_\mu \gamma_5 {\vec \tau} \cdot (\partial^\mu {\vec \phi}) 
+ \frac{g}{2} \gamma_\mu {\vec \lambda} \cdot {\vec A}^\mu  \right] \psi \, \theta_V  , \label{int-lag}
\ene
with ${\vec \lambda}$ the SU(3) generators and $g$ the quark-gluon coupling constant. 
The free pion field and the kinetic energy of the gluon field, ${\vec A}^\mu$, are, respectively, described by ${\cal L}_{\pi}$ and 
${\cal L}_{g}$. 

We firstly calculate the second-order energy correction to the nucleon or delta mass.  
The energy shift of a multi-quark, ground state, $|0 \rangle$, due to the interaction is given by 
the Hubbard's prescription 
\bge
E-E_0 = \langle 0 | \sum_{m=1}^\infty (-i)^m \frac{1}{m!} \int i\delta(t_1) d^4x_1 \cdots \int d^4x_m 
T[{\cal H}_{int}(x_1) \cdots {\cal H}_{int}(x_m)] | 0 \rangle_{con.} ,  \label{hubbard}
\ene
where ${\cal H}_{int}$ is the interaction Hamiltonian density.  
The energy shift is then given as $E^{(2)} = E_{dr} + E_{nd}$, 
where the first term is the direct contribution and the second one is the non-direct contribution. 
(See Eqs.(\ref{direct}) and (\ref{nondirect}) later.)

The noninteracting, quark green function is given by 
$iG^0(r,r^\prime) = \langle 0 | T[\psi(r){\bar \psi}(r^\prime)] | 0 \rangle$, and it can be separated into two pieces: 
$G^0(r,r^\prime) = \int \frac{d\omega}{2\pi} e^{-i\omega(t-t^\prime)} 
[G^0_F({\vec r},{\vec r\,}^\prime, \omega) + G^0_D({\vec r},{\vec r\,}^\prime, \omega)]$. 
The first term is the usual Feynman propagator in a spherical cavity (bag) and 
the second one describes the occupied, multi-quark ground state~\cite{chin} 
\bge
G^0_D({\vec r},{\vec r\,}^\prime, \omega) = \sum_{n\leq n_F} 
U_n({\vec r}){\bar U}_n({\vec r\,}^\prime) 2\pi i \delta(\omega - E_n) ,  \label{qgreenD}
\ene
where $U_n$ is the positive energy state with a complete set of quark 
quantum numbers $n (=\{ \nu \kappa \mu \mu_i \mu_c \})$ including isospin $\mu_i$ and color $\mu_c$ 
($n_F$ specifies the quantum numbers at the Fermi surface in a hadron). 

Here we restrict the expansion of the quark propagator to the ground state, 
i.e., $\nu =0$ and $\kappa = -1$. 
Such a truncation may be considered as a regularization of the 
quark propagator, where in flavor SU(2) the intermediate baryon states in loop diagrams are restricted to 
the nucleon and delta~\cite{inoue}. 
This is consistent with the idea of the CBM. Thus, we let $n$ label the spin, isospin and color $\{\mu \mu_i \mu_c\}$. 

The pion propagator is defined by 
$i\Delta_{ab}(r,r^\prime) = \langle 0 | T[\phi_a(r) \phi_b(r^\prime)] | 0 \rangle = i \delta_{ab} 
\Delta(r,r^\prime)$, where $(a, b)$ specifies the isospin. It is then given by the multipole expansion 
\bge
\Delta(r,r^\prime) 
= \sum_{\ell, m} \int \frac{d\omega}{2\pi} e^{-i\omega(t-t^\prime)} 
\Delta_\ell(r,r^\prime, \omega) Y_{\ell m}({\hat r}) Y_{\ell m}^*({\hat r}^\prime) . \label{pgreen} 
\ene
%
%where $\Delta_\ell(r,r^\prime, \omega)$ is the radial part of the propagator. 

The gluon propagator can be calculated in the Coulomb gauge\footnote{
It can be shown that the result does not depend on the choice of the gauge~\cite{inoue}.} 
\bg
iD_{00}^{cd}(r,r^\prime) &=& i \delta_{cd} \int \frac{d^4k}{(2\pi)^4} 
\frac{e^{-i k\cdot(x-x^\prime)}}{{\vec k}^2} = i \delta_{cd} D_{00}(r,r^\prime) ,  \label{ggreen0} \\
iD_{ij}^{cd}(r,r^\prime) &=& i \delta_{cd} \int \frac{d^4k}{(2\pi)^4} 
\frac{e^{-i k\cdot(x-x^\prime)}}{{\vec k}^2 + i\epsilon} \left( \delta_{ij} - \frac{k_i k_j}{{\vec k}^2} \right) 
= i \delta_{cd} D_{ij}(r,r^\prime) ,  \label{ggreenT} 
\en
where $D_{00}^{cd}$ ($D_{ij}^{cd}$) represents the Coulomb (transverse) propagator, $(i,j) = 1, 2, 3$ and 
$(c, d)$ specifies the color. 

Using these propagators, the direct (or Hartree) contribution between two quarks is calculated by
\bge
E_{dr} = - \frac{1}{2} \int \!\!\! \int \delta(t_1) d^4r_1 d^4r_2 \, \Theta(r_1,r_2) \, 
{\rm tr}[\Gamma_1 {\vec t}_1\, G^0_D(r_1,r_1)] \cdot {\rm tr}[\Gamma_2 {\vec t}_2\, G^0_D(r_2,r_2)],  \label{direct}
\ene
where $\Gamma_{1, 2}$ represents the vertex (ps, pv or gluon) including the coupling constant 
(the Lorentz index is suppressed here), $\Theta$ stands for the pion or gluon propagator 
and ${\vec t} = {\vec \tau}/2$ or ${\vec \lambda}/2$.  The non-direct contribution is given by
\bge
E_{nd} = \frac{1}{2} \int \!\!\! \int \delta(t_1) d^4r_1 d^4r_2 \, \Theta(r_1,r_2) \, 
{\rm tr}[\Gamma_1 {\vec t}_1\, G^0(r_1,r_2) \cdot \Gamma_2 {\vec t}_2\, G^0(r_2,r_1)].  \label{nondirect}
\ene
This consists of the exchange contribution, $E_{ex}$, which is evaluated with $G^0=G^0_D$ in Eq.(\ref{nondirect}), and 
the self-energy contribution, $E_{qsf}$, where the pion is emitted and absorbed by the same quark. 
$E_{qsf}$ is calculated by replacing one of $G^0_D$'s with $G^0_F$ in $E_{ex}$.  
The case with simultaneously $G^0=G^0_F$ is removed, because it is the vacuum energy. 
The Hartree-Fock (HF) contribution between two quarks is defined by $E_{qHF} = E_{dr} + E_{ex}$.  
Note that these energy shifts correspond to the case where a closure approximation to the intermediate states is taken 
in the CBM, and that such an approximation is reasonable. 

For example, the HF result for the pv coupling, $E_{qHF}^{pv}$, is given by
\bg
E_{qHF}^{pv} &=& 
-\frac{[{\vec \sigma} {\vec t}\, ]_{HF} N^4}{12\pi^2 x^3 (f_\pi R)^2 R}   
\int_0^x d\rho_1 \rho_1^2 \int_0^x d\rho_2 \rho_2^2 \int_0^\infty \frac{dt \, t^4}{t^2+y^2} 
\biggl[ A(\rho_1) A(\rho_2) j_0(t\rho_1) j_0(t\rho_2)  \nonumber \\
&+& \frac{1}{3} 
\left\{ A(\rho_1) B(\rho_2) j_0(t\rho_1) \left( j_0(t\rho_2) -2 j_2(t\rho_2) \right) 
+ A(\rho_2) B(\rho_1) j_0(t\rho_2) \left( j_0(t\rho_1) -2 j_2(t\rho_1) \right) \right\} \nonumber \\
&+& \frac{1}{9} B(\rho_1) B(\rho_2) \left( j_0(t\rho_1) -2 j_2(t\rho_1) \right) 
\left( j_0(t\rho_2) -2 j_2(t\rho_2) \right) \biggr] ,  \label{hfpv1}
\en
with $N$ the normalization constant for the quark wave function, $R$ the bag radius, 
$x$ the lowest quark eigenvalue, $y=m_\pi R/x$, 
$A(\rho) = j_0^2(\rho) - \beta^2 j_1^2(\rho)$, $B(\rho) = 2 \beta^2 j_1^2(\rho)$, 
$\beta = x/(\alpha+\delta)$, $\alpha^2 = x^2 + \delta^2$ and $\delta = mR$. 
Here the spin-isospin matrix element is given by~\cite{chin}
\bge
[{\vec \sigma} {\vec t}\, ]_{HF} = 
\sum_{i \neq i^\prime\in N, \Delta} \langle i| {\vec \sigma} {\vec t} |i^\prime \rangle \cdot \langle i^\prime| {\vec \sigma} 
{\vec t} |i \rangle 
= 9-S(S+1) - I(I+1) ,  \label{me1}
\ene
where the index ($i, i^\prime$) runs over the spin and isospin, and 
$S \, (I)$ is the total spin (isospin) of N or $\Delta$. 
Because the intermediate baryon states are restricted to the lowest mode, the self-energy contribution, $E_{qsf}^{pv}$, has 
the same form as the HF result except for the spin-isospin matrix element: $[{\vec \sigma} {\vec t}\, ]_{sf} = 27/4$ for both N and 
$\Delta$. 

The pion-induced baryon self-energies should reproduce the correct, leading non-analytic (LNA) behavior of chiral 
perturbation theory ($\chi$PT)~\cite{lna}. 
The LNA contribution is associated with the infrared behavior of the baryon self-energy, and hence, for example, 
Eq.(\ref{hfpv1}) gives $-3g_A^2m_\pi^3/(32\pi f_\pi^2) \times (30/25, 6/25)$ for the (N, $\Delta$) in the infrared limit. 
In contrast, $E^{pv}_{qsf}$ gives $-3g_A^2m_\pi^3/(32\pi f_\pi^2) \times 27/25$ for both N and $\Delta$. 
Thus, the total amount is $-3g_A^2m_\pi^3/(32\pi f_\pi^2) \times (57/25, 33/25)$ for the (N, $\Delta$), which is 
precisely the leading-order correction given by large $N_c$ $\chi$PT. 
However, because of the closure approximation 
taken for the intermediate states, the term of $m_\pi^4 \ln(m_\pi)$ does not appear in the present calculation. 

The contribution from the ps (gluon) interaction, $E_{qHF, qsf}^{ps} (E_{qHF, qsf}^{g})$, is also calculated 
in the similar manner.\footnote{
The energy shift due to the 
Coulomb propagator vanishes because of the color charge neutrality~\cite{mit}.
}
Note that there exists a nonvanishing, interference (sv) contribution between the ps and pv couplings, $E_{qHF, qsf}^{sv}$.  
%Numerically, the ps coupling is very small, because it is ${\cal O}(m^2/f_\pi^2)$. 
%The main energy shift comes from the pv coupling and its 
%magnitude is ${\cal O}(1/f_\pi^2 R^2)$. 
The pv, ps and gluon corrections lower the baryon mass, while 
the interference contribution increases it but its magnitude is ${\cal O}(m/f_\pi^2R)$ and thus small. 

Each correction is a function of the bag radius $R$, and, for example, $E_{qHF}^{pv}$ 
diverges like $ \sim - 1/R^3$ as $R \to 0$.  Thus, the bag collapses as 
$R \to 0$. 
%This can, however, be cured by introducing a non-local interaction between the quark and 
%pion. 
%The pion has a distinguishable property from the other hadrons, i.e., it is the only particle whose 
%geometrical size is smaller than its Compton wave length.  So, the elementary field 
%description of the pion field may be adequate at low energy (large scale).  However, 
Because the pion has a finite size, 
the effect of the $q{\bar q}$ substructure is essentially important when the bag 
radius is very small. 
In Ref.~\cite{saito}, a phenomenological, non-local interaction was studied 
to settle this collapse at $R \sim 0$. 
The effect of the $q{\bar q}$ substructure of pion can eventually be described by a form factor at the vertex of the 
quark-pion interaction. When the charge radius of the pion is about $0.56$ fm~\cite{dally}, 
the form factor is estimated as
\bge
F_{q\pi}(R) = \frac{1}{\left( 1+1.3\times (b / R)^2 \right)^{3/2}} ,  \label{f.f.}
\ene
with $b = 0.46$ fm (see Fig.2 in Ref.\cite{saito}).  
%Note that $F_{q\pi} \to R^3$ as $R \to 0$.  
%Multiplying the energy corrections due to the pion cloud by $F_{q\pi}^2(R)$, 
Using this form factor, one can get the finite, pion-loop contributions.  

The coupling between a quark and gluon is scale-dependent and the lowest-order coupling at 
momentum transfer $Q^2$ is 
$\alpha_s(Q^2) = g^2/4\pi = 12\pi/[(33-2N_f) \ln (Q^2/\Lambda_{QCD}^2)]$ 
with $N_f$ quark flavors and $\Lambda_{QCD} \simeq 200$ MeV.  
%As $Q \to \Lambda_{QCD}$, the coupling diverges, signaling the onset of confinement. 
%In the present model, because this effect has already been taken into account in the 
%bag model, we should take $\alpha_s(Q^2)$ to saturates at some critical value $\alpha_s^c$ as $Q \to 0$~\cite{capstick}. 
In practice, this behavior can be parametrized in a convenient form~\cite{capstick}
\bge
\alpha_s(Q^2) = \sum_{k} \alpha_k e^{-Q^2/4\gamma_k^2} 
= a_1 \, e^{-4Q^2} + 0.25\, e^{-Q^2} + 0.15 \, e^{-Q^2/10} 
+ 0.2 \, e^{-Q^2/1000} ,  \label{alphac5}
\ene
where $Q^2$ in GeV$^2$ and the parameters, $\alpha_k$ and $\gamma_k$, except $a_1$ are constrained to follow the 
behavior of $\alpha_s(Q^2)$. We treat $a_1$ as a parameter. 
The form (\ref{alphac5}) is convenient, because it is 
easily transformed into the form in coordinate space 
\bge
\alpha_s(R) = \sum_{k} \frac{2\alpha_k}{\sqrt{\pi}} \int_0^{\gamma_k R} e^{-x^2} dx .  \label{alphac4}
\ene
Here we assume that this gives the coupling constant at the scale of the bag radius, $R$, and that 
the energy shift due to the gluon exchange is given by replacing $g^2$ with $4\pi \alpha_s(R)$ in $E_{qHF, qsf}^{g}(R)$.  
Note that $\alpha_s(R) \to 0$ as $R \to 0$. %and that it weakens the gluon contribution near $R\sim 0$. 

Now we are in a position to present the numerical result for 
the N or $\Delta$ mass in free space.  The mass is given by a sum of the usual 
bag energy~\cite{mit} and the corrections due to the pion and gluon exchanges. 
We fix the current quark mass $m = 5$ MeV, because the dependence of the baryon mass on $m$ is very weak. 
There are four parameters: $B$, $z_N$, $z_\Delta$ and $a_1$ (in $\alpha_s$).  
Since we can expect that the usual $z$ parameter for the N is not much different from that for the $\Delta$, we 
choose $z_0 = z_N = z_\Delta$. Then, the bag constant, $B$, and $z_0$ are determined so as to fit 
the free nucleon mass, $M_N (= 939$ MeV), 
with its radius $R_N = 0.6$ or $0.8$ fm. 
The remaining parameter, $a_1$, is fixed so as to yield
the correct mass difference between $M_N$ and $M_\Delta (=1232$ MeV)
together with the pion-cloud contribution. 
We then find $B^{1/4} = 231.8 \, (183.7)$ MeV, $z_0 = 2.46 \, (1.17)$ and $a_1 = 5.01 \, (6.08)$ for $R_N = 0.6 \, (0.8)$ fm.

\begin{figure}
\includegraphics[scale=0.9]{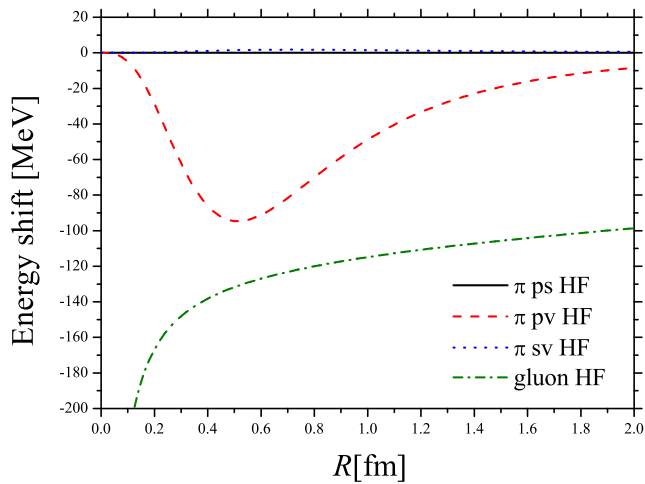}%[width=15cm,height=7cm,clip]
\caption{\label{fig:pion} 
Energy shift due to the HF contribution ($R_N=0.8$ fm). 
The dot-dashed curve presents the gluon contribution, which approaches a constant as $R \to 0$. 
The solid (dashed) [dotted] curve is for the ps (pv) [sv] contribution from the pion exchange.  
}
\end{figure}
\begin{table}[h]
\caption{\label{tab:energyshift} Energy shift (in MeV) due to the pion or gluon exchange.}
\begin{ruledtabular}
\begin{tabular}{cccccc}
 & $R$(fm) & $E_{qHF}^{pv+ps+sv}$ & $E_{qsf}^{pv+ps+sv}$ & $E_{qHF}^g$ & $E_{qsf}^g$ \\
\hline
N        & 0.6  & -89.5 & -80.5 & -111.1 & -333.3 \\
$\Delta$ & 0.611 & -17.7 & -79.7 & 110.7 & -332.0 \\
\hline
N        & 0.8  & -68.3 & -61.5 & -120.0 & -360.1 \\
$\Delta$ & 0.823 & -13.7 & -59.1 & 119.4 & -358.1 \\
\end{tabular}
\end{ruledtabular}
\end{table}
In Table~\ref{tab:energyshift}, we present the energy corrections. 
The N-$\Delta$ mass difference mainly comes from 
the gluon-exchange HF contribution. 
Note that the N-$\Delta$ mass difference due to the pion cloud is 
about $60$ MeV (for $R_N = 0.8$ fm), and that is near the upper limit 
allowed from lattice QCD constraints~\cite{young}.  
%
%\begin{figure}
%\includegraphics[scale=1.2]{vertex}%[width=15cm,height=7cm,clip]
%\caption{\label{fig:vertex} 
%Vertex corrections to the $\sigma$-quark and $\omega$-quark interactions. }
%\end{figure}
%
In Fig.~\ref{fig:pion}, for example, we show the HF energy due to the pion or gluon exchange 
as a function of the bag radius. As expected, the ps contribution is quite small. 
The interference is also small and its sign is positive. Because of Eqs.(\ref{f.f.}) and (\ref{alphac4}), 
the energy shift is finite everywhere and thus the total energy for the N or $\Delta$ mass has (global) one minimum at a certain $R$. 

To describe a nuclear matter, we need the intermediate attractive and short-range repulsive nuclear forces. 
As the QMC model is based on the one-boson-exchange (OBE) picture~\cite{qmc}, it is achieved by introducing 
the $\sigma$ and $\omega$ mesons.\footnote{
It should be noticed that the OBE model is still the most economical and quantitative 
phenomenology for describing the nuclear force~\cite{mach}.
} 
However, the present $\sigma$ meson is chirally singlet and {\em not} the chiral partner of the $\pi$ meson.  
This $\sigma$ represents, in some way, the exchange of two pions in the iso-scalar channel~\cite{delorme}. 

Now let us start from the following Lagrangian density for 
the ``chiral quark-meson coupling (CQMC) model'': 
${\cal L}_{CQMC} = {\cal L}_{CBM} + {\cal L}_{\sigma \omega}$, 
where 
\bge
{\cal L}_{\sigma \omega} = {\bar \psi} \left[ g_\sigma^q \sigma - g_\omega^q \gamma_0 \omega \right] \psi \, \theta_V  
- \frac{1}{2} m_\sigma^2 \sigma^2 + \frac{1}{2} m_\omega^2 \omega^2 ,  \label{qmc-lag2}
\ene
with $g_\sigma^q (g_\omega^q)$ the $\sigma (\omega)$-quark coupling constant, $m_\sigma (m_\omega)$ the meson mass 
and $\sigma (\omega)$ the mean-field value of the $\sigma$ ($\omega$) meson. 
%Because the quark is dressed by the pion cloud, the pion-loop diagrams may contribute to the interaction vertices 
%(see Fig.~\ref{fig:vertex}).  When we take the mean-field approximation for the $\sigma$ and $\omega$ mesons, 
%such effect can be, however, absorbed into the coupling constant (at zero momentum transfer). 

In an iso-symmetric nuclear matter, the total energy per nucleon is given by
\bge
E_{tot} = \frac{4}{(2\pi)^3 \rho_B} \int^{k_F} d{\bf k} \sqrt{{\bf k}^2+M_N^{*2}} + 3g_\omega^q \omega 
+ \frac{1}{2}( m_\sigma^2 \sigma^2 - m_\omega^2 \omega^2 ) ,  \label{tot}
\ene
with $\rho_B = 2k_F^3/3\pi^2$ and $M_N^{*}$ the effective nucleon mass. 
The attractive force due to the $\sigma$ changes the quark mass in matter as 
$m^* = m - g_\sigma^q \sigma$ ($m^*$ the effective quark mass), which modifies the quark wave function. 
This modification generates the effective nucleon mass, $M_N^{*}$, in matter. 
Because the change of the quark wave function varies the source of the $\sigma$ field, 
we have to solve the coupled, nonlinear equations for the nuclear matter self-consistently 
(for details, see Refs.\cite{qmc}).  
%Note that, depending on $\rho_B$, the $\pi$-N coupling also varies self-consistently through $F_{q\pi}(R)$. 

%
\begin{table}[h]
\caption{\label{tab:cqmc} Coupling constants and calculated properties for symmetric nuclear matter at $\rho_0$. 
The last three columns show the relative changes (from their values at zero density) of the bag radius, the lowest eigenvalue 
and the root-mean-square (rms) radius of the nucleon calculated with 
the quark wave function. 
The nucleon mass and the nuclear incompressibility, $K$, are in MeV.}
\begin{ruledtabular}
\begin{tabular}{cccccccc}
$R_N$(fm) & $g_\sigma^2/4\pi$ & $g_\omega^2/4\pi$ & $M_N^*$ & $K$ & $\delta R_N^*/R_N$ & $\delta x^*/x$ & $\delta r^*/r$ \\
\hline
0.6 & 6.11 & 10.20 & 632 & 362 & 0.01 & -0.18 & 0.05 \\
0.8 & 4.93 & 9.59 & 647 & 365 & 0.02 & -0.22 & 0.07 \\
\end{tabular}
\end{ruledtabular}
\end{table}
\begin{figure}
\includegraphics[scale=1.0]{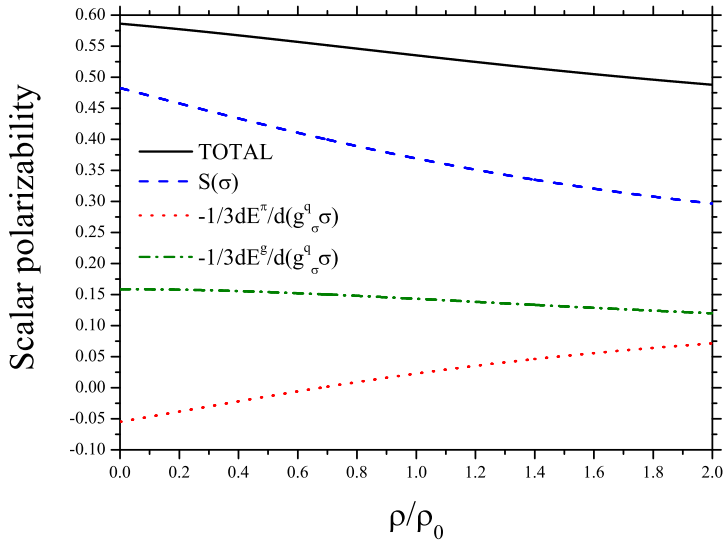}%[width=15cm,height=7cm,clip]
\caption{\label{fig:scalarpol} 
Scalar polarizability in the CQMC model (for $R_N=0.8$ fm). 
The dashed curve presents the scalar polarizability in the usual QMC model, while the solid one is for the CQMC model. 
The dotted (dot-dashed) curve shows the pion (gluon) contribution to the scalar polarizability. 
}
\end{figure}
The numerical result for the nuclear matter 
(with $m_\sigma = 550$ MeV, $m_\omega = 783$ MeV and $R_N = 0.6$ or $0.8$ fm) is presented in Table~\ref{tab:cqmc}. 
The $\sigma$-N and $\omega$-N coupling constants, 
$g_\sigma (= 3g_\sigma^q S(\sigma=0))$ and $g_\omega (= 3g_\omega^q)$, are determined to fit 
the nuclear saturation condition ($-15.7$ MeV) at normal nuclear density $\rho_0 (= 0.15$ fm$^{-3}$). 
Here $S(\sigma)$ is the scalar density calculated by the quark wave function~\cite{qmc}. 

In the CQMC model, the bag radius is swelled by a few percent at $\rho_0$. 
The quark eigenvalue, $x$, decreases by about $20$\%, which leads to 
the smaller in-medium nucleon mass than in the QMC model. Although the rms radius of a nucleon swells by about 
$6$\% at $\rho_0$, it may still be within the experimental constraint~\cite{electron}.  

In Fig.~\ref{fig:scalarpol}, we present the scalar polarizability in the CQMC model, which is given by a sum of 
the quark scalar density, $S(\sigma)$, and the contributions from the pion and gluon exchanges. 
We find that even in the CQMC model the scalar polarizability decreases with increasing $\rho_B$. 
Because of this reduction of the scalar polarizability in matter, the present model can achieve 
the nuclear saturation property with a much smaller value of the nuclear incompressibility, $K$, than in QHD.  
%As in the QMC model, the CQMC model can also describe the nuclear matter well. 

%
\begin{figure}
\includegraphics[scale=0.9]{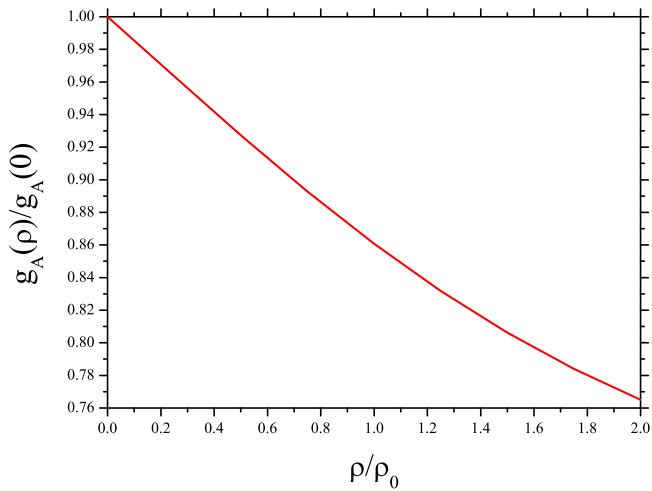}%[width=15cm,height=7cm,clip]
\caption{\label{fig:ga} 
Variation of $g_A$ in matter (for $R_N=0.8$ fm). }
\end{figure}
In Fig.~\ref{fig:ga}, as an example, we show the axial vector coupling constant, $g_A$, in matter. 
If the pion is massless, the pion pole term 
in the axial vector current gives just $1/3$ of 
the quark core contribution 
i.e., the MIT-bag-model value~\cite{chin,tsushima}. 
However, when the pion is massive, 
the pion current does not contribute to $g_A$. 
Here the quark core contribution is thus calculated with the corrections 
due to the Peierls-Thouless projection method 
and the Lorentz contraction of the bag~\cite{tsushima}.  We obtain $g_A = 1.15$ in free space. 
The CQMC predicts a reduction of about $14$\% for $g_A$ at $\rho_0$. 
Note that the pion field surrounding a nucleon becomes 
also weaker with increasing $\rho_B$ since 
the pion itself is mainly generated by the pv interaction. 

In the present model, of importance is the scalar polarizability. 
It describes the response of a quark to the scalar field in matter 
and leads to the reduction of the $\sigma$-N coupling constant~\cite{qmc,formf}.  Such response arises from the change of the quark 
wave function in a medium, and it is thus the many-body effect~\cite{skyrme}. 
Then, the CQMC predicts that the iso-scalar, central force is weakened depending on $\rho_B$. 
On the other hand, in EFT, the (supposed) $\sigma$ exchange in the OBE model can be well understood by 
the correlated, two-pion contribution~\cite{kaiser1,donoghue}. 
However, recent calculations for matter show that, 
the N$^3$LO potential based on EFT produces 
very deep overbinding at large $\rho_B$~\cite{li}.  
Thus, this reduction of the central force may be favorable. 
To draw more definite conclusions, however, further studies are necessary.

%The iso-scalar, central potential is then expressed in terms of $g_A$, $c_1$ and $c_3$, where 
%$c_i$ ($i=1 \sim 4$) are the LECs in the next-to-leading order term of the effective chiral $\pi N$ lagrangian~\cite{kaiser1}. 
%The LEC $c_3$ is also related to the nucleon axial polarizability~\cite{ericson}.
%Furthermore, 
%the $c_1$ part of the central potential is proportional to the nucleon scalar form factor (or the $q{\bar q}$ condensate), 
%which may also vary in matter~\cite{sigmaterm}.  The LEC $c_3$ is related to the so-called nucleon axial polarizability, 
%while $c_4$ is given in terms of the nucleon isovector magnetic form factor~\cite{electron}. 
%Thus, the present CQMC approach is very promising to predict how the LECs vary depending on $\rho_B$. 
%For more practical purposes, 
%it is possible to 
%construct a point-coupling model of nuclear matter based on in-medium chiral perturbation theory, in which the 
%nucleon self-energies are expanded in terms of $\rho_B$ up to ${\cal O}(k_F^5)$.  The coefficients appearing in the expansion are 
%then determined so as to fit the results of Dirac-Brueckner calculations.  The QMC may also be able to explain 
%why the coefficients are such values. 

In summary, we have developed for the first time a chiral version of 
the quark-meson coupling model based on the cloudy bag model, in which 
the effects of pion cloud and gluon exchange are included self-consistently.
The model can describe a symmetric nuclear matter reasonably well. 
We have also shown that $g_A$ decreases with increasing $\rho_B$. 
This implies that the iso-scalar, central nuclear force 
should be weakened in matter. 
At ${\cal O}(1/f_\pi^2)$, the CBM Lagrangian automatically 
provides the Weinberg-Tomozawa (WT) term, 
%${\cal L}_{WT} = - (1/4f_\pi^2) {\bar \psi} \gamma_\mu \tau \psi ({\vec \phi} \times \partial^\mu {\vec \phi}) \theta_V$,  
which is a new source of $g_A$~\cite{morgan,tsushima}. 
Thus, in the future, it is desirable to perform a self-consistent calculation 
up to ${\cal O}(1/f_\pi^2)$ including the WT term.
%the magnitude of the WT correction is proportional to the product of the upper and lower components of the 
%quark wave function~\cite{morgan} and it thus increases as the density grows.  
%However, $g_A$ eventually decreases in matter 
%because the WT contribution is expected to be smaller than the quark core contribution. 
%Using the CQMC model, it is very intriguing and the challenge to precisely calculate the density dependence of 
%the LECs, $g_A$ etc. appearing in EFT. 
It is also very interesting to compare the CQMC with the EFT approach 
to investigate the internal 
structure change of the nucleon in medium. 

\begin{acknowledgements}
The authors thank A.W. Thomas for valuable discussions on the pion-cloud effect. 
This work was supported by Academic Frontier Project (Holcs, Tokyo University of Science, 2005) of MEXT, 
and by the US Department of Energy, Office of Nuclear Physics, through contract no. DE-AC05-06OR23177, 
under which Jefferson Science Associates, LLC, operates Jefferson Lab.
\end{acknowledgements}
%

%\clearpage
%
%%%%%%%%%%%%% Bibliography %%%%%%%%%%%%%%%%%%%%%%%%%%%%%%%%%
%
%\newpage

%

%\newpage %Just because of unusual number of tables stacked at end

\end{document}